# THE DIGITAL DIVIDE IN GERIATRIC CARE: WHY USABILITY, NOT ACCESS, IS THE REAL PROBLEM

By: Christine Ine

**Abstract**

The rapid increase in the world's aging population to 16% by the year 2050 spurs the need for the application of digital health solutions to enhance older individuals' independence, accessibility, and well-being. While digital health technologies such as telemedicine, wearables, and mobile health applications can transform geriatric care, their adoption among older individuals is not evenly distributed. This study redefines the "*digital divide*" among older health care as a *usability divide,* contends that user experience (UX) poor design is the primary adoption barrier, rather than access. Drawing on interdisciplinary studies and design paradigms, the research identifies the main challenges: visual, cognitive, and motor impairment; complicated interfaces; and lack of co-creation with older adults, and outlines how participatory, user-focused, and inclusive notions of design can transcend them. Findings reveal that older persons easily embrace those technologies that are intuitive, accessible, and socially embedded as they promote autonomy, confidence, and equity in health. The study identifies the effects of the design attributes of high-contrast screens, lower interaction flow, multimodal feedback, and caregiver integration as having strong influences on usability outcomes. In addition, it critiques the current accessibility guidelines as being technically oriented rather than experiential and demands an ethical, empathetic understanding of design grounded in human-centered usability rather than technical accessibility in itself. Thus, we contend that closing the usability gap in gerontechnology mandates a paradigm shift—away from designing for old people to designing with them. It recommends the use of co-design, emotional engagement, and long-term support systems in digital health innovation to position technology as an advocate of autonomy, equity, and dignity in old age.

**Keywords:** Digital health technologies, older adults, usability and user experience (UX),

telemedicine and mobile health, inclusive design

**1.0     Introduction**

The aging population of the world is growing exponentially from 10% in 2022 to an estimated 16% in 2050, which is driving the growth of long-term and chronic diabetes, cardiovascular disease, and neurodegenerative disorders (Boersma et al., 2020). Multimorbidity in older people drives healthcare use and costs up, and while letting down autonomy and quality of life down (Volders et al., 2019). Despite improvements in the provision of healthcare, access for this group remains limited by sociodemographic, economic, and health-related factors such as mobility impairment, low income, amongst others. Digital health technologies, ranging from telehealth platforms to wearables and mobile health apps, have emerged as signature technologies to increase access to care and promote autonomy for elderly populations (Bertolazzi et al., 2024). This has created gerontechnology, designed to enable "*ageing in place*" with constant, home-based, and low-cost care. However, older adults remain the least represented active users of digital health tools. The prevailing storyline attributes this "*digital divide*" to older adults' inability or unwillingness to learn new technologies. Evidence showed the opposite: when technologies are inexpensive, intuitive, and socially supported, older adults embrace them with enthusiasm (Alsswey et al., 2023; Lee et al., 2024). The actual barrier is not access but usability. Cluttered interfaces full of jargon, ambiguous workflows, confusing workflows, and non-intuitive navigation frustrate patients, erode trust, and validate disengagement.

The majority of studies and interventions focus on increasing access through the provision of internet access, devices, or digital literacy training without fixing design problems that make digital health technologies irrelevant in practice (Czaja et al., 2021). It is against the aforementioned backdrop, this study reframes the digital divide in geriatric health care as a usability divide, arguing that poor user experience (UX) design, rather than restricted access, is the greatest hindrance to digital health adoption for older persons. By a review of user-centric design, participatory development, and inclusive UX testing methodologies, this study demonstrates how well-designed digital health technology can encourage autonomy, confidence, and equitable healthcare for the elderly.

## 2.0     Barriers and Facilitators of Technology Adoption in Older Adults

The current exponential growth rate of the ageing population worldwide is expected to surpass 2.1 billion by 2050 (WHO, 2022), necessitating economically viable models of healthcare, of which telemedicine is at the forefront. Telemedicine debunks spatial and temporal disparities in care, offering remote consultations, round-the-clock monitoring, and better management of chronic conditions (Narasimha et al., 2017). It benefits included increased efficiency, reduced costs, and more robust patient–provider relationships. Despite these, adoption is inconsistent among older adults. Contrary to prevailing stereotypes, most seniors are receptive to digital health technologies, provided they are convenient, accessible, and integrated into well-understood systems of care. The barriers are context and design, and not necessarily age. Low digital literacy, poor usability, technological unfamiliarity, and socioeconomic disadvantage all contribute to limiting adoption. Complex navigation, fonts too small to read, and jargon deter use, and this underlines the value of designs that take into account physical (vision, dexterity) and cognitive (memory, attention) constraints (Alsswey et al., 2023). The MOLD-US model characterizes such barriers as four dimensions (i.e., perception, cognition, motivation, and physical ability) emphasizing vision, memory, and fine motor impairments as key constraints (Gomez-Hernandez et al., 2023).

**Fig. 1:** MOLD-US Framework of Barriers to Digital Health Adoption in Older Adults

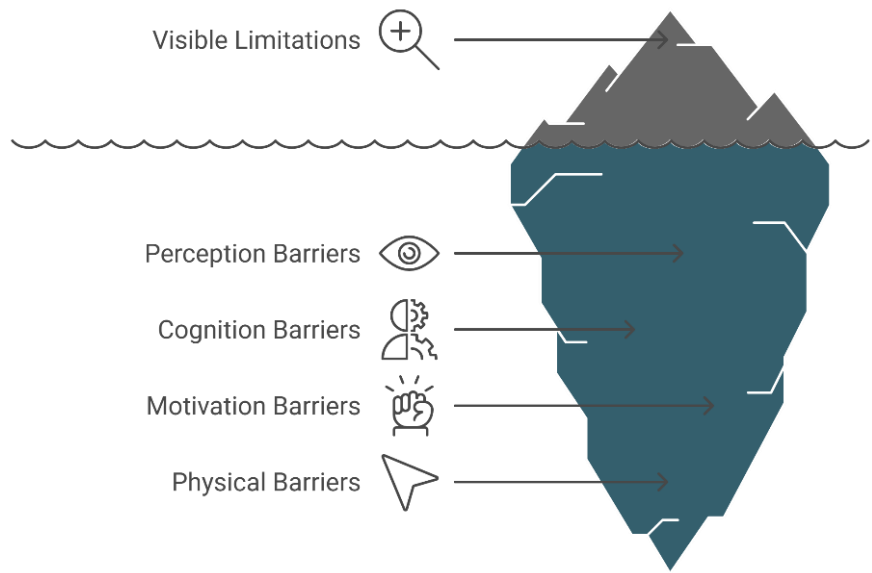

Source: Author's Visual (2025).

Similarly, the Technology Acceptance Model (TAM) relates perceived usefulness and ease of use to user acceptance. Interface complexity decreases usefulness, but ease of use increases satisfaction and adoption. Empirical studies have repeatedly indicated that usability, cost savings, and fit to context are the strongest predictors of adoption. Trust and confidence are built on usability, affordability, a history of good experience, and carer engagement (Lee et al., 2024; Bertolazzi et al., 2024). Strong social relationships, especially where carers are more technologically adept, facilitate exploration and extended use (Şahin et al., 2024). Likewise, interfaces with familiar care providers and infrastructures permit continuity and familiarity (Odebunmi et al., 2024). Where technology reduces travel and increases availability, demand for face-to-face contact, technical avoidability, and exclusionary design persist (Ilali et al., 2023). Education, wealth, and cultural backgrounds can also act as mediators for adoption processes (Mao et al., 2022; Knotnerus et al., 2024). Bridging the usability gap is a matter of redesigning digital health for older users. Technologies must:

i. Be co-designed with seniors and caregivers to accommodate real needs.

ii. Employ clear visual hierarchies, large fonts, and plain language.

iii. Supply flexible interaction modes (voice, touch, gesture) to accommodate around impairments.

iv. Blend with current healthcare routines to enhance trust.

v. Offer ongoing support and iterative feedback mechanisms.

Where such principles guide development, digital health technologies become enabling, not displacing, enabling older people to sustain autonomy and connection in later life.

## 3.0 Broader Tech Trends and Accessibility Standards

Technology adoption among older adults is growing fast. A recent AARP national survey found that most adults 50+ now use technology to enhance daily living, independence, and well-being (Kakulla, 2024). Two-thirds of respondents said digital tools, especially telemedicine, health apps, and safety monitoring systems, improve their lives. Caregivers are also using GPS tracking, motion sensors, and emergency response systems to support care. But despite the enthusiasm, technological spending is constrained by cost and usability issues, not disinterest or inability. The problem is the mismatch between mainstream product design and age-related limitations like visual, auditory, and motor decline. Many older users struggle with inaccessible layouts, small buttons, and information-dense screens. This gap between capability and design usability has ethical and practical implications: it perpetuates exclusion in systems meant to promote independence.

Current responses to this issue include international accessibility standards like the Web Content Accessibility Guidelines (WCAG) developed by the W3C's Web Accessibility Initiative (WAI), which already address most aging-related needs (Arch et al., 2024). But compliance often produces *technically accessible but functionally frustrating* systems. To close this gap, recent studies propose a holistic usability model that goes beyond basic accessibility. He et al. (2025) outline four design domains—functional framework, interaction logic, visual design, and user experience. These emphasize intuitive navigation, responsive interfaces, clear visual hierarchy, and emotional engagement. As Rogers et al. (2023) reported, sustained adoption among older adults is not just about whether technology is "*usable*" but whether it feels intuitive, empowering, and emotionally supportive. So, aging-friendly design must move from minimal accessibility compliance to human-centered usability standards that address emotional satisfaction and trust.

## 4.0   User-Centered and Inclusive Design Approaches

With the global population aging rapidly, user-centered design is a moral and practical necessity in healthcare technology. Electronic personal health records (ePHRs), for example, can help with chronic disease monitoring and patient engagement. But usability problems – complex workflows, cognitive overload, and poor customization – continue to hold back adoption (Zhang & Song, 2024). The gap is that many design frameworks treat "*old age*" as a single condition rather than a diverse experience. Conventional approaches assume chronological age to be the only factor determining digital capacity, ignoring motivation, self-efficacy, and cultural context. Solutions are hence technically correct but experientially incorrect.

To address this, more recent research is focusing on co-creation and participatory design where older adults are involved directly in the design process. Participatory design encompasses experiential nuances, frustration points, language use preferences, and contextual routines that designers would otherwise not notice. As Stamate et al. (2024) show, optimism and curiosity drive adoption, while low confidence and negative attitudes block it. Thus, design must target psychological empowerment as much as physical accessibility. Inclusion is also complicated by cultural and social factors. The different backgrounds of older adults lead to different degrees of comfort with giving and receiving feedback, and experimenting; thus, the inclusive design should be responsive to the different cultures. However, most eHealth systems are still fragmented and quite poorly interoperable, which results in a lack of trust even in the case of perfect technical design (Aiesha, 2021). On the other hand, the answer is not more technology but rather more inclusion. The systems should be participatory, context-sensitive, and psychologically enabling. This means the designers will have to treat the elderly as partners and co-creators instead of just the recipients. When such a change occurs, technology is then the means to facilitate health equality, dignity, and continuous involvement, rather than an obstacle to care.

## 5.0     UX Design Preferences and Guidelines for Older Adults

Digital health technologies have become a part of the aging process in a way that they cannot be separated, but their success is still dependent on a good design that takes into consideration the physical, cognitive, and perceptual realities of elderly users. The fact that an elderly person goes through a decline in vision, memory, and digital literacy does not necessarily mean that he or she is not willing to use technology; the real issue, in fact, is the usability of the current interfaces. There are consistent findings across empirical research that very small and evidence-based design changes lead to large increases in engagement. According to Zhang and Song (2024), aged people show a preference for the use of high-contrast colors, skeuomorphic design, and icon-based menus—these characteristics enhance readability and reduce the need for text. They also indicate that the use of simple grid layouts, mobile-first interfaces, and audio cues can be quite beneficial for users with decreased memory and vision. In the same vein, Li (2022) and Liu et al. (2021) reported that a lot of frustration and mistakes with the use of chronic disease management apps come from cluttered screens, unclear data presentation, and long interaction steps. Easier visualization and stepwise prompting reduce cognitive load significantly. Yaldaie (2022) similarly determines that subtle visual and structural enhancements (larger buttons, sans-serif fonts, and less abrupt interface transitions) significantly improve usability for dementia patients and older adults as well. The findings determine three basic UX objectives:

i. Maximize visual clarity with high-contrast colors and sparse layouts.

ii. Reduce cognitive load with rational, responsive sequences of interactions.

iii. Engage multimodal cues (visual, auditory, tactile) to facilitate understanding.

These principles extend beyond usability enhancement; they foster autonomy, confidence, and sustained digital interaction. Figure 2 below displays the underlying UX design strategies that enhance accessibility and inclusion for older adult users.

**Fig. 2:** UX Design Strategies for Older Adult Engagement

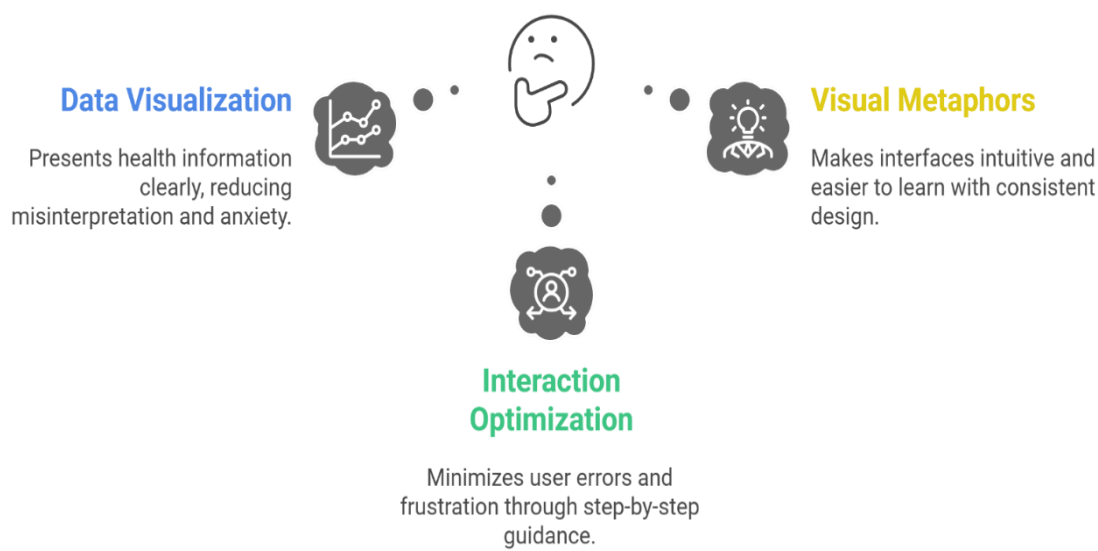

Source: Author's Visual (2025)

## 6.0 UX Evaluation, Usability Testing, and System Design in Practice

As technologies assume an increasing role in the provision of care for older adults, usability emerges as a central factor in successful adoption. While age-related declines in cognition, vision, and mobility are well documented, the real issue is how systems are designed, tested, and refined to address these declines. Current evidence shows that participatory and iterative approaches yield the best outcomes. Studies of home-monitoring and exergaming interventions reveal that co-designed systems—those refined with older users' feedback—demonstrate superior usability, engagement, and safety (Villalba-Mora et al., 2021; Chu et al., 2021). Similarly, incorporating older adults' mobility and social needs into interface guidelines reduces barriers and strengthens independence (Bahrampoor et al., 2025). However, existing evaluation practices still fall short. The System Usability Scale (SUS), although widely used, prefers standard scoring and does not include emotional and contextual use variables. More complex measures like the User Experience Questionnaire (UEQ) and models like the Technology Acceptance Model (TAM) provide insight but remain yet to be adopted (Qiu et al., 2019). Remote usability testing, both synchronous and asynchronous, also faces the challenge of accommodating the comfort and feedback patterns of older adults (Takano et al., 2023; Kondratova et al., 2023).

System-specific studies also validate chronic flaws: hard navigation, ambiguous headings, and low contrast text weaken motivation and confidence (Srinivas et al., 2017; Siette et al., 2023). Technological flaws contribute to blanket accessibility limitations such as inadequate font readability and the absence of voice-help interfaces (Zainal et al., 2023). Ad-hoc usability issues can be fixed through iterative redesign, but sustained long-term use demands ongoing support, motivation, and training. Gerontechnology usability thus extends beyond interface cosmetics. It is based on participatory, long-term, and context-aware design and evaluation. By placing emphasis on flexibility, simplicity, and multimodal interaction, digital health systems can enhance not only usability but also confidence, autonomy, and inclusion of older adults.

## 7.0 Conclusion and Recommendations

This study dispels the historic illusion that older adults are immune to technology. Evidence shows that older adults learn, use, and maintain the use of digital health technologies if the system is usable, accessible, and socially supported (Czaja et al., 2021; Siette et al., 2023). The greatest obstacle is not accessing rather usability, overly complex, graphically dense, or cognitively demanding interfaces (Narasimha et al., 2017; Li, 2022; Zhang & Song, 2024). Also, it is established that accessible user-centered design is the most robust predictor of long-term uptake in gerontechnology (Ottaviani et al., 2022; Yaldaie, 2022). In an attempt to promote equitable digital health uptake, the following are proposed recommendations:

i. **Emphasize User-Centered Design**

Engage older adults at every phase of design—needs discovery, prototyping, and testing. Co-creation builds trust, contextually relevant, and experience-based usability.

ii. **Implement Accessibility and Usability Guidelines**

In addition to WCAG compliance, incorporate age-adjusted functionalities like large text, high contrast colors, tactile or auditory feedback, and simple iconography.

iii. **Streamline Interaction Flows**

Design sequential, logically structured interactions with lower cognitive demands. This enhances accuracy, confidence, and self-efficacy among older users.

iv. **Enhance Social and Caregiver Integration**

Define caregivers as enablers of technology through mutual dashboards, remote monitoring, and collaborative communication features.

### v. Enable Policy and Industry Alignment

Governments and industry stakeholders for digital health innovation should include inclusion, affordability, and interoperability by governments and industry leaders.

Therefore, bridging the digital divide in elderly care requires a shift from access-centered interventions to usability-centered design. The technology must be designed empathetically, collaboratively, and with sensitivity towards human diversity so that it becomes a facilitator of autonomy, equity, and dignity for the aged, rather than a divider.